\documentclass[twocolumn,aps,prb,showpacs,amsmath,superscriptaddress]{revtex4}
\usepackage{amssymb}
\usepackage{mathrsfs}
\usepackage{graphicx}
\usepackage{float}

\begin{document}
\title{Characterization of symmetry-protected topological phases in polymerized models by trajectories of Majorana stars}
\author{Chao Yang}
\affiliation{Beijing National Laboratory for Condensed
Matter Physics, Institute of Physics, Chinese Academy of Sciences, Beijing 100190,
China}
\author{Huaiming Guo}
\affiliation{Department of Physics, Beihang University, Beijing, 100191, China}

\author{Li-Bin Fu}
\affiliation{National Laboratory of Science and Technology on Computational Physics,
Institute of Applied Physics and Computational Mathematics, Beijing 100088, China}

\author{Shu Chen}
\thanks{schen@aphy.iphy.ac.cn}
\affiliation{Beijing National Laboratory for Condensed
Matter Physics, Institute of Physics, Chinese Academy of Sciences, Beijing 100190,
China}
\affiliation{Collaborative Innovation Center of Quantum Matter, Beijing, China}
\date{ \today}

\begin{abstract}
By using Majorana's stellar representation, we give a clear geometrical interpretation of the topological phases of inversion-symmetric polymerized models by mapping the Bloch states of multi-band systems to Majorana stars on the Bloch sphere. While trajectories of Majorana stars of a filled Bloch band exhibit quite different geometrical structures for topologically trivial and nontrivial phases, we further demonstrate that these structures are uniquely determined by distributions of Majorana stars of two high-symmetrical momentum states, which have different parities for topologically different states.
\end{abstract}
\pacs{
03.65.Vf, 
73.21.Cd, 
03.65.Aa, 
05.30.Fk 
}
\maketitle

\section{Introduction}
Recently, Majorana's stellar representation (MSR) has gained revived interest
as it offers an intuitive geometrical approach to understanding quantum systems with multiple components by mapping states in the higher
dimensional space as points on a Bloch sphere \cite{Majorana1932,Schwinger,Hannay,Niu2012,Bruno2012,Fu2014}. In general, a quantum state of spin-1/2 systems or equivalently two-level systems can
be represented as a point on a unit sphere \cite{Bloch}, and the evolution of the quantum state corresponds to a particular trajectory
on the Bloch sphere. By using MSR, which represents a quantum pure state of spin-J systems in terms of a symmetrized state of 2J spin-1/2 systems, one can generalize this geometric approach to large spin systems or multi-level systems. In terms of MSR, the evolution of a spin-$J$ state can be intuitively understood by  trajectories of $2J $ points on the two-dimensional (2D) Bloch sphere, with these $2J$ points generally coined as Majorana
stars (MSs). This approach naturally provides an intuitive way to study
high spin systems from geometrical perspectives, which has made the MSR a useful tool
in many different fields, e.g.,
classification of entanglement in symmetric quantum states \cite{N-qubit,Mandilara,Ganczarek,Ribeiro,Markham}, analyzing the spectrum of the Lipkin-Meshkov-Glick  model \cite{Vidal}, studying Bose condensate with high spins \cite{Demler,Lamacraft,Zhaihui,Ueda}, and calculating geometrical phases of large-spin systems \cite{Bruno2012,Fu2014}.

As much of the search for the application of MSR has focused on high-spin systems, it is interesting to apply this approach to study the multi-band topological systems. It is well known that a topological insulator distinguishes a trivial band insulator by its nontrivial topological energy band, which has different geometrical property from a trivial band \cite{TI-RMP}. For a two-band system, e.g., the Su-Schrieffer-Heeger (SSH) model \cite{SSH}, one can map the Bloch state into a 2D Bloch sphere, and the geometrical meaning of topologically different phases can be unveiled by their distinct trajectories \cite{Montambaux,Ryu}. As a paradigmatic topological model \cite{Shen-book}, the SSH model supports either topologically trivial or nontrivial phase, characterized by the quantized Berry phase $0$ or $\pi$ \cite{Berry,Zak,Niu-RMP}, which is experimentally measurable as demonstrated in the recent cold atom experiment \cite{cold1}.
In this work, we shall apply MSR to study one-dimensional (1D) topologically nontrivial polymerized systems, which can be viewed as a  multi-band generalization of the celebrated dimerized SSH model \cite{SSH,SS} and were also proposed to be realizable in optical superlattice systems \cite{Guo,Ganeshan}. The application of MSR enables us to investigate the geometrical property of muti-band topological systems by mapping the multi-level quantum states to MSs on the Bloch sphere. Consequently, a filled Bloch band forms specific trajectory of MSs on the Bloch sphere, which exhibits very different geometrical structure for topological or trivial phase. Furthermore, we unveil that the distinct geometrical structures of trajectories for topologically different states are closely related to their parities and determined by the distribution of MSs at two high-symmetry momentum points.
Our study provides an intuitive way to distinguish topologically different phases of multi-band systems and unveils the intrinsic relation between the band topology and parity from the geometrical point of view, which shall deepen our understanding of their geometrical properties.

\section{1D polymerized model with inversion symmetry}
We consider a 1D superlattice with period $T$ $(T\ge2)$ described by the Hamiltonian:
\begin{eqnarray}
H = \sum_{n} \sum_{\alpha=1}^{T}(t_{\alpha} \hat{c}_{\alpha, n}^{\dagger} \hat{c}_{\alpha+1,n}+h.c.) ,
\end{eqnarray}
where $\hat{c}_{T+1,n} \equiv \hat{c}_{1,n+1}$ and we require the system having inversion symmetry by enforcing
\begin{equation}
t_{\alpha}=t_{T-\alpha} .
\end{equation}
For the case with $T=2$, it reduces to the SSH model \cite{SSH}. Models with $T \geq 3$ can be viewed as generalizations of the dimerized SSH model \cite{SS}, and for convenience, we also refer them as polymerized models. As the unit cell consists of $T$ sites, the spectrum of the system splits into $T$ bands. By taking a Fourier transformation
$\hat{c}_{\alpha,n} = 1/\sqrt{N} \sum_k e^{ikn} \hat{c}_{\alpha k}$ with $\alpha=1,2,\cdots,T$,  the Hamiltonian in the momentum space can be written as
$H =\sum_{k}\psi_{k}^{\dagger}h(k)\psi_{k}$ with $\psi_{k}=(c_{1 k},c_{2k},\cdots,c_{Tk})^{T}$ and
\begin{equation}\label{Ham}
h(k)=\left(
  \begin{array}{cccccc}
    0 & t_{1} & 0 & \cdots & 0 & t_{T}e^{-ik} \\
    t_{1} & 0 & t_{2} & \cdots & 0 & 0 \\
    0 & t_{2} & 0 & \cdots & 0 & 0 \\
    \vdots & \vdots & \vdots & \ddots & \vdots & \vdots \\
    0 & 0 & 0 & \cdots & 0 & t_{T-1} \\
    t_{T}e^{ik} & 0 & 0 & \cdots & t_{T-1} & 0 \\
  \end{array}
\right) .
\end{equation}
For the general case with arbitrary $t_\alpha$, the system does not have topologically nontrivial properties. Nevertheless, when the system has inversion symmetry, it may support topologically nontrivial phase characterized by the quantized nontrivial Berry phase or the emergence of doubly degenerate edge states for the open chain \cite{Guo}. For convenience, we shall focus our study on a concrete case with $T=3$ in the present work, and also show our results can be directly generalized to cases with larger periods, e.g., the case with $T=4$.

For the $T=3$ superlattice model or trimerized model with inversion symmetry, the hopping amplitudes can be parameterized as
\begin{equation}
t_1=t_2=t(1-\delta), ~~ t_3=t(1+\delta), \label{trimer}
\end{equation}
where $t=1$ is taken as the unit of energy and $|\delta|<1$ is set. The spectrum is split into $3$ bands and gap between bands is always open for any nonzero $\delta$. Considering the state with the lowest band being fully filled, we find that the state is topologically different for $\delta>0$ and $\delta<0$. For the open chain with the length $L=3N$, there appear degenerate edge states at both ends for $\delta >0$ but none for $\delta <0$. This result suggests that there exists a topological phase transition by varying $\delta$ with the transition point at $\delta=0$. For the periodic chain, the topological phase transition can be characterized by the change of the Berry phase of the system, i.e., the Berry phase $\gamma = \pi$ in the topological phase and  $\gamma = 0$ in the trivial phase. Here the Berry phase across the Brillouin zone (BZ), also known as Zak phase \cite{Zak}, is defined as $\gamma=i\oint\langle \Phi(k)|\frac{d}{dk}|\Phi(k)\rangle dk$ with $\Phi(k)$ denoting the occupied Bloch states.

The existence of topological states in our polymerized models is protected by inversion symmetry \cite{Guo,inversion}. In the momentum space, the inversion symmetry means that $\hat{P} h (k)\hat{P}^{-1} = h (-k)$, where the inversion operator $\hat{P}$
is an anti-diagonal matrix with the the matrix element given by $P_{i,j}=\delta_{i,T+1-j}$.
The system also has time reversal symmetry, which leads to $\hat{T}h(k)\hat{T}^{-1}=h(-k)$, here the time reversal operator $\hat{T}$ is just the complex-conjugation operator $\hat{K}$. As we shall see in the following context, both the inversion and time reversal symmetries give some restrictions on the MSR of Bloch states, which plays an important role in determining trajectories of MSs of topologically different states.

\section{Majorana Representation and Berry Phase}
For a $T$-band system, the Bloch state can be expressed as $ |\Phi(k)\rangle = \sum_{\alpha=1}^{T} C_\alpha (k) | \alpha \rangle_k $. To represent this multi-level state by MSs, it is convenient to map the state to a spin-J state $ |\Phi(k)\rangle = \sum_{m= -J}^{J} C_m (k) |J, m \rangle $ with $J=(T-1)/2$.
There is a one-to-one correspondence between parameters $C_\alpha (k)$ and $C_m (k)$ by taking $m=\alpha-1-J$.
According to Schwinger boson representation theory \cite{Schwinger}, the angular momentum operators can be described by creation and annihilation operators of two mode bosons, $\hat{a}^{+}$, $\hat{a}$, and $\hat{b}^{+}$, $\hat{b}$, and the state $|J, m \rangle $  can be expanded by
$
|J,m \rangle = [(J+m)!(J-m)!]^{-1/2} (\hat{a}^{+})^{J+m}(\hat{b}^{+})^{J-m}|\o\rangle
$,
where $|\o\rangle$ is defined by $\hat{a}^{+}|\o\rangle=|\uparrow\rangle$ and $\hat{b}^{+}|\o\rangle=|\downarrow\rangle$.
With the help of Schwinger representation, the state $\Phi(k)$ can be factorized as
\begin{equation}\label{Phi}
\begin{split}
|\Phi(k) \rangle =&\frac{1}{N_{J}}\prod_{j=1}^{2J}(\cos{\frac{\theta_{j}}{2}}\hat{a}^{+}+\sin{\frac{\theta_{j}}{2}}e^{i\phi_{j}}\hat{b}^{+})|\o \rangle
\end{split}
\end{equation}
where $N_{J}$ is the normalization coefficient.
If we denote $x_{j}=\tan{\frac{\theta_{j}}{2}}e^{i\phi_{j}}$, then the factorization parameters $\theta_{j}$ and $\phi_{j}$ can be determined by  the roots of the following polynomial equation \cite{Majorana1932}
\begin{equation}\label{poly}
\sum_{j=0}^{2J}\frac{(-1)^{j}C_{J-j}(k)}{\sqrt{(2J-j)!j!}}x^{2J-j}=0.
\end{equation}
From Eq.(\ref{Phi}), it is obvious that $|\Phi(k) \rangle$ can be viewed as the product of $2J$ spin-$\frac{1}{2}$ states with
$|u_{j} \rangle =\left(
             \cos{\frac{\theta_{j}}{2}},
             \sin{\frac{\theta_{j}}{2}}e^{i\phi_{j}} \right)^{T}
$,
and a given $T$-band Bloch state can be described by $T-1$ MSs on the Bloch sphere.

Considering the time reversal symmetry and inversion symmetry of the inversion-invariant polymerized model, we can directly get the eigenstate $|\Phi \rangle$ fulfilling the following relations:
\begin{eqnarray}
|\Phi^{*}(k) \rangle &=&  |\Phi(-k) \rangle ,\label{eq1} \\
|\Phi^{*}(k) \rangle &=& \hat{P} |\Phi(k) \rangle . \label{eq2}
\end{eqnarray}
From Eq.(\ref{eq1}), we get $C^{*}_{j}(k) =C_{j}(-k)$.
So if $x(k)=\tan{\frac{\theta}{2}} e^{i \phi}$ is the solution of Eq.(\ref{poly}),
then $x^{*}(k)$ must be the solution of
$
\sum_{j=0}^{2J}( -1)^{j}C_{J-j}(-k)/\sqrt{(2J-j)!j!} {x^*}^{2J-j}=0 .
$
In other words, for any star $\vec{u}_{j}(k)$ of $(\theta_{j}(k), \phi_{j}(k))$, there must exist a star $\vec{u}_{l}(-k)$ satisfied $\theta_{j}(k)=\theta_{l}(-k)$, $\phi_{j}(k)=-\phi_{l}(-k)$. So the whole trajectory of MSs for a filled Bloch band is symmetric about the meridian of the Bloch sphere.
Similarly, from Eq.(\ref{eq2}), we have $C_{J-j}(k)= C^{*}_{j-J}(k)$, $j=0,1,\cdots,n$.
So if $x=\tan{\frac{\theta}{2}} e^{i\phi}$ is a solution of Eq.(\ref{poly}), $x^{'}=(\frac{1}{x})^{*}=\tan{\frac{\pi-\theta}{2}} e^{i\phi}$ is also a solution. That is to say, for a fixed momentum k, any star $\vec{u}_{j}$ of $(\theta_{j}, \phi_{j})$ corresponds to a star $\vec{u}_{l}$ of $(\theta_{l}=\pi-\theta_{j}, \phi_{l}=\phi_{j})$, unless $\vec{u}_{j}$ lies on the equator. So the whole trajectory of MSs for a filled band is always symmetric about the equator.
\begin{figure}[htbp]
  \centering
  \includegraphics[width=1.0\linewidth]{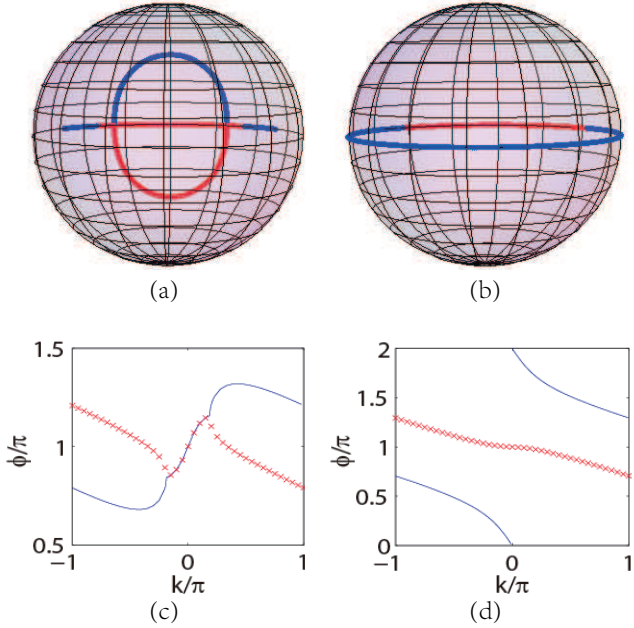}\\
  \caption{Trajectories of MSs for the lowest Bloch band of the trimerized model with inversion symmetry. The blue and red curves represent different MSs. (a) topologically trivial state with $\delta=-0.2$, the Berry phase is $\gamma_{1}=0.404$, $\gamma_{2}=-0.404$, $\gamma_{c}=0$. (b) topologically nontrivial state with $\delta=0.2$, the Berry phase is $\gamma_{1}=-2.216$, $\gamma_{2}=-0.925$, $\gamma_{c}=0$. (c) and (d) show the change of $\phi$ of MSs for the lowest Bloch band of trimerized model as $k$ goes from $-\pi$ to $\pi$, corresponding to (a) and (b), respectively.
  }\label{fig1}
\end{figure}

In terms of the MSR, the Berry phase for the $T$-band polymerized model can be
represented as a sum of two parts, $\gamma=\gamma_{0}+\gamma_{c}$, where
\[
\gamma_{0}=\sum_{j} \gamma_{j}=-\sum_{j}\frac
{1}{2} \oint(1-\cos{\theta_{j}})d\phi_{j},
\]
and the phase
\[
\gamma_{c}= \frac{1}{2}\sum_{i<j}^{2J} \oint \frac{\vec{u}_{i}\times\vec{u}_{j}\bullet d(\vec{u}_{j}-\vec{u}_{i})}{N_{J}^{2}}\frac{\partial N_{J}^{2}}{\partial d_{ij}}
\]
represents correlations between any two MSs \cite{Fu2014},
where $\vec{u}_{i}$ is a function of $k$ and the integral is carried out along the evolution paths of MSs as the momentum goes over the BZ.
By using symmetries of MSs, we can prove that $\gamma_c = 0 $ as the summation terms of correlations cancel out each other (see the appendix for details). Consequently, the Berry phase of the occupied Bloch band is simplified to the sum of Berry phases of MSs on the Bloch sphere, generated by the evolution of each star. Since the MSs either appear in pairs, in terms of $\vec{u}_{j}$ and $\vec{u}_{l}(\theta_{l}=\pi-\theta_{j},\phi_{l}=\phi_{j})$ for $\theta_{j} \ne\frac{\pi}{2}$, or locate in the equator, we can further simplify the expression of Berry phase by using $\sum_j \cos \theta_j =0$. Finally, the Berry phase of an occupied Bloch band is simplified to
\begin{equation}\label{phase}
\gamma = -\sum_{j}\frac{1}{2} \oint d\phi_{j},
\end{equation}
which is the sum of integrals of each MS along its projecting trace on the equator.

For the trimerized model parameterized by Eq.(\ref{trimer}), the eigenstate $|\Phi(k) \rangle $ can be described by two MSs. As $k$ goes across the BZ, trajectories of the lowest Bloch band for cases with $\delta<0$ and $\delta>0$ are shown in Fig.\ref{fig1}(a) and (b), respectively. It is obvious that trajectories for topologically different states display distinct geometric structures. Here, in the topologically trivial phase, the Berry phase of each star cancels out. However, in the topologically nontrivial phase, the trajectories of two stars splice together to cover the whole equator, which gives $\gamma=\pi$.
To see this more clearly, we show the change of $\phi_i$  as a a function of momentum in Fig.\ref{fig1}(c) and (d), corresponding to  Fig.\ref{fig1}(a) and (b), respectively. For the topologically trivial case,  as $k$ goes from $-\pi$ to $\pi$, the integral over $\phi$ for each star cancels out each other, which generates a zero Zak phase.

\begin{figure}[htbp]
  \centering
  \includegraphics[width=0.8\linewidth]{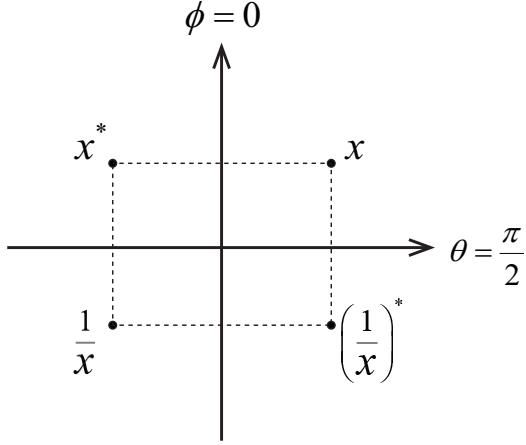}\\
  \caption{Schematic plot for the distribution of Majorana stars of high-symmetry momentum state. At the high-symmetry point, every four stars form a group which is mirror-symmetrical about both the meridian and equator.} \label{pro2}
\end{figure}

Next we show that different topological structures of trajectories of MSs are closely related to properties of MSs at $k=0$ and $k=\pi$. At these high-symmetry points denoted by $k_s$, which fulfils $k_s=-k_s$ up to a reciprocal lattice vector, we have $\hat{T} h (k_s)\hat{T}^{-1} = h (k_s)$ and $\hat{P} h (k_s)\hat{P}^{-1} = h (k_s)$.  These symmetries suggest that the distribution of MSs for the high-symmetry state $\Phi(k_s)$  must be mirror-symmetrical about both the meridian and equator.  The state $\Phi(k_s)$ also has a certain parity, i.e., $\hat{P}|\Phi(k_s) \rangle=\xi(k_s) |\Phi(k_s) \rangle $ with $\xi(k_s) = \pm 1$, and the parity $\xi$ of a filled Bloch band is given by the product of $\xi(0)$ and $\xi(\pi)$.
Using the expression of $\hat{P}$, we get
\begin{gather}\label{app21}
  C_{J-j}(k_{s}) = \xi(k_{s})C_{j-J}(k_{s}), \\\label{app22}
  \xi(k_{s})e^{i\sum_{j}\phi_{j}} = 1 ,
\end{gather}
where all coefficients $C_{j}(k_{s})$ are real due to the time-reversal symmetry. Substituting (\ref{app21}) into the Eq.(6), we find that if $x_{j}$ is the solution of the equation, then $x_{j}^{*}$, $\frac{1}{x_{j}}$ and $(\frac{1}{x_{j}})^{*}$ are also the solutions, i.e., the Majorana stars for the high-symmetry state are mirror-symmetrical about both the meridian and equator, as schematically shown in Fig. \ref{pro2}. If $\theta_{j}=\frac{\pi}{2}$, the four stars are degenerated to 2 stars on the equator; if $\phi_{j}=0$ or $\pi$, the stars are degenerated to two stars on the meridian; and if $\theta_{j}=\frac{\pi}{2}$ and $\phi_{j}=0$ or $\pi$, all the stars are degenerated to a single star.
For the odd-parity state, $\xi(k_{s})=-1$, and we get $\sum_{j}\phi_{j}=\pi$ from Eq. (\ref{app22}). While for the even-parity state, $\xi(k_{s})=1$, and we get $\sum_{j}\phi_{j}=0$.

\begin{figure}[htbp]
  \centering
  \includegraphics[width=0.8\linewidth]{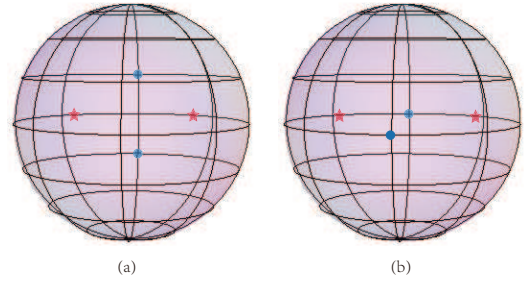}\\
  \caption{MSR for the high-symmetrical momentum state of the trimerized model with (a) $\delta=-0.2$ and (b)  $\delta=0.2$, respectively. Blue points and red stars correspond to the state at $k=0$ and $k=\pi$, respectively. While the parity for the point of $k=\pi$ is always even, the parity for the point of $k=0$ can be even for the trivial phase (a) or odd  for the topological phase (b). } \label{pot}
\end{figure}

For the trimerized model with $T=3$, there are only two Majorana stars corresponding to a high-symmetry momentum state. For the odd-parity state, the stars must be located at $\vec{u}_{1}(\frac{\pi}{2},0)$ and $\vec{u}_{2}(\frac{\pi}{2},\pi)$ in order to fulfil the mirror-symmetry conditions and the condition of $\phi_{1}+\phi_{2}=\pi$ simultaneously. While for the even-parity state, the stars can be located at $\vec{u}_{1}(\frac{\pi}{2},\phi)$ and $\vec{u}_{2}(\frac{\pi}{2},-\phi)$  or at $\vec{u}_{1}(\theta,0)$ and $\vec{u}_{2}(\pi-\theta,0)$, or at $\vec{u}_{1}(\theta,\pi)$ and $\vec{u}_{2}(\pi-\theta,\pi)$.
To see it clearly, we display the distribution of MSs at the high-symmetry points for both the topologically trivial (in Fig.\ref{pot}(a)) and nontrivial phases (in Fig.\ref{pot}(b)). While the parity of $\Phi(\pi) $ is always even with $\xi(\pi)=1$ for both phases, we have $\xi(0)=1$ for the trivial phase, and $\xi(0)=-1$ for the topological phase.
\begin{figure}[htbp]
  \centering
  \includegraphics[width=1.0\linewidth]{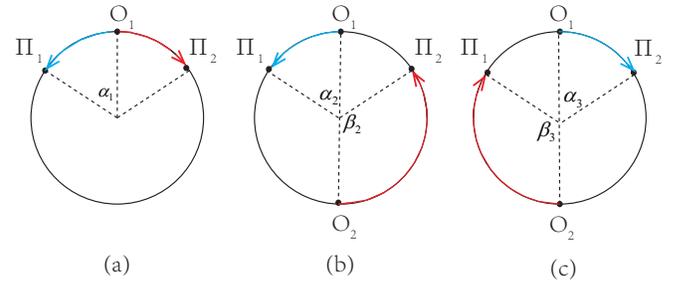}\\
  \caption{Schematic plot for the change of polar angle of MSs when $k$ goes from $0$ to $\pi$, (a) for topologically trivial phase, (b) and (c) for topologically nontrivial phase. Here $O_{j}$ correspond to MSs at $k=0$ and $\Pi_{j}$ to MSs at $k=\pi$.}\label{exp}
\end{figure}

As the topologically trivial and nontrivial phases have different parities, next we unveil the Berry phase of the corresponding Bloch band can be uniquely determined by distributions of MSs at two high-symmetry points. To calculate the Berry phase, it is convenient to project the trajectory of MSs on the Bloch sphere to the equator, according to Eq.(\ref{phase}). Consequently the MSs symmetrically located above and below the equator merge to the same point. Then, we divide the BZ into two parts $(0,\pi)$ and $(\pi,2\pi)$, and we just need to consider the interval $(0,\pi)$ as the integral of $\phi$ over the interval $(\pi,2\pi)$ (equivalently $(-\pi,0)$) gives the same contribution to the Berry phase as over the interval $(0,\pi)$, according to the symmetry analysis. Hence, the Zak phase $\gamma$ is just the double of the Berry phase as $k$ integrates from $0$ to $\pi$.
In the topologically trivial phase, when $k=0$, both MSs are located at ``$O_1$" on the projected equator shown in Fig.\ref{exp}(a). As $k$ travels from $0$ to $\pi$, the stars move to $\Pi_{1}$ and $\Pi_{2}$, generating changes of polar angles $\Delta \phi_{1}=2m\pi+\alpha_{1}$ and $\Delta \phi_{2}=2n\pi-\alpha_{1}$, respectively, where $m, n$ can be any integer. From Eq.(\ref{phase}), we get the Zak phase $\gamma=-(\Delta \phi_{1}+\Delta \phi_{2})=-2(m+n)\pi$, i.e., $\gamma\mod(2\pi)=0$. In the topologically nontrivial phase, MSs corresponding to the $k=0$ state are located at ``$O_1$" and ``$O_2$", and there are two possibilities for the evolution of state from $k=0$ to $k=\pi$, as shown in Fig.\ref{exp}(b) and (c). For the case of Fig.\ref{exp}(b), $O_{1}$ moves to $\Pi_{1}$ and $O_{2}$ to $\Pi_{2}$ as $k$ goes from $0$ to $\pi$, with corresponding changes of polar angles given by $\Delta\phi_{1}=2m\pi+\alpha_{2}$ and $\Delta\phi_{2}=2n\pi+\beta_{2}$. Since $\alpha_{2}+\beta_{2}=\pi$, we get the Zak phase $\gamma=-2(m+n)\pi-\pi$, i.e., $\gamma\mod(2\pi)=\pi$. For the case of Fig.\ref{exp}(c), we can get the same conclusion by following similar discussions.

Our results can be directly generalized to general polymerized models with inversion symmetry. Concretely,
we consider a tetramerized model with $t_1=t_3=t$, $t_2=t(1-\delta)$ and $t_4=t(1+\delta)$, and take $t=1$ and $|\delta|<1$.
Similar to the case of $T=3$, there exists a topological phase transition occurring at $\delta=0$, with $\delta<0$ corresponding to the trivial phase and $\delta>0$ the topologically nontrivial phase. The analysis of geometric meaning is analogous to the trimerized model despite a little more complicated.
\begin{figure}[htbp]
  \centering
  \includegraphics[width=1.0\linewidth]{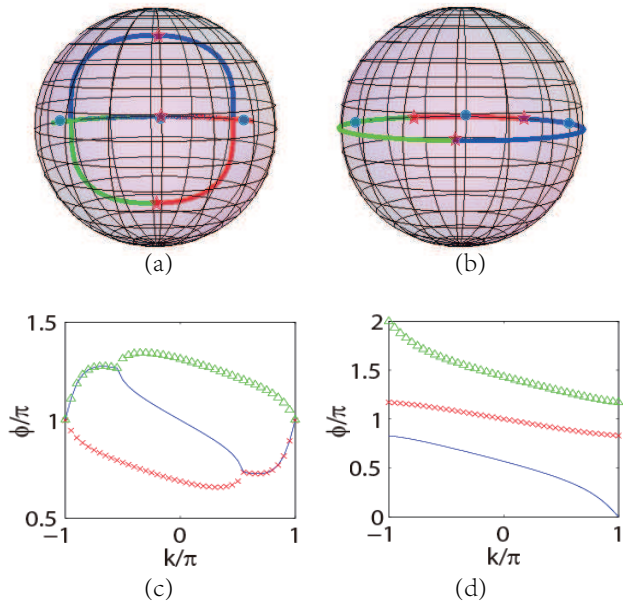}\\
  \caption{MSR for the topologically trivial and nontrivial state of the tetramerized model with (a) $\delta=-0.2$ and (b)  $\delta=0.2$, respectively. The blue, red and green curves represent different MSs. Blue points and red stars represent MSs of states at $k=0$ and $k=\pi$, respectively.
  (c) and (d) display the change of $\phi$ of MSs as $k$ goes from $-\pi$ to $\pi$, corresponding to (a) and (b), respectively.
  }\label{fig4}
\end{figure}

For the tetramerized model with $T=4$, each state in the momentum space is represented by three MSs, and their trajectories over the BZ are shown in Fig.\ref{fig4}(a) and (b) for topologically trivial and nontrivial phases, respectively. MSs in two high-symmetry points are also marked on the trajectories. In the topologically trivial phase, all trajectories of MSs go back and forth, whereas in the topologically nontrivial phase, the trajectories form a close circle and cover the equator completely. Consequently, Zak phases for the topologically trivial and nontrivial states are $0$ and $\pi$, corresponding to Fig.\ref{fig4}(c) and (d), respectively. Similarly, different MS distributions at $k=\pi$ for topologically trivial and nontrivial states indicate that they have different parities.

\section{Summary}
In summary, we have unveiled the geometrical properties of topological phases of 1D inversion-invariant multi-band systems from  trajectories of MSs, which exhibit different topological structures for topologically different phases. By utilizing the symmetric properties of MSs, we found that the Zak phase of a filled band can be represented as the summation of integral of each MS along its projecting trace on the equator, and takes $0$ and $\pi$ for topologically trivial and nontrivial phases, respectively. We further demonstrated that the topological structure of the trajectory of MSs is closely related to the parity of the system, which is determined by properties of Bloch states at two high-symmetry points.

\begin{acknowledgments}
S. C. is supported by NSFC under Grants No. 11425419, No. 11374354 and No. 11174360 and by National Program for Basic Research of MOST. H. G. is supported by NSFC
under Grants No.11274032 and No. 11104189. L. B. F. is supported by the
NSFC under Grants No. 11374040.
\end{acknowledgments}

\appendix
\section{The calculation of correlation phase}
\begin{figure}[htbp]
  \centering
  \includegraphics[width=0.7\linewidth]{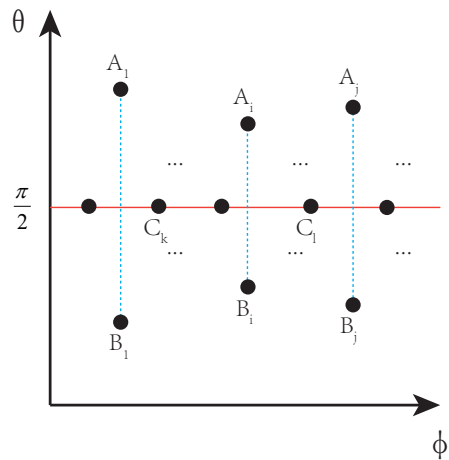}\\
  \caption{Schematic plot for the distribution of all the Majorana stars on the $\theta-\phi$ plane for a Bloch state with fixed $k$. The stars are classified into three classes denoted by $A_{i}$, $B_{i}$ and $C_{i}$, respectively.} \label{sfig1}
\end{figure}

The correlation phase $\gamma_c$ in the main text can be represented as $\gamma_c =  \oint d\gamma_{c}$ with
\begin{equation}
d\gamma_{c}=\frac{1}{2}\sum_{i<j}^{2J}\frac{\vec{u}_{i}\times\vec{u}_{j}\bullet d(\vec{u}_{j}-\vec{u}_{i})}{N_{J}^{2}}\frac{\partial N_{J}^{2}}{\partial d_{ij}},
\end{equation}
where the normalization coefficient $N_{J}$ is a symmetric function for permuting any two stars $\vec{u}_{i}$ and $\vec{u}_{j}$, and $d_{ij}=1-\vec{u}_{i}\bullet\vec{u}_{j}$ is the distance.
It is clear that $\gamma_c$  is a sum of correlations between any two stars.
Due to the existence of time reversal symmetry and inversion symmetry, the Majorana stars for a Bloch state with fixed momentum $k$ distribute either on the equator or symmetrically about the equator. So the distribution of Majorana stars for a Bloch state can be schematically displayed in Fig. \ref{sfig1}, where $A_{i}$ are the stars above the equator, $B_{i}$ are the corresponding ones below the equator symmetrical to $A_{i}$, and $C_{i}$ are on the equator.

Firstly, we calculate the correlation between $A_{i}$ and $A_{j}$ and the corresponding correlation between $B_{i}$ and $B_{j}$, which are given by
\begin{eqnarray}
& & d\gamma_{cA_{i}A_{j}}   \nonumber \\
&=& \frac{dk}{2}
  \left|\begin{array}{ccc}
    \sin{\theta_{A_{i}}}\cos{\phi_{A_{i}}} & \sin{\theta_{A_{i}}}\sin{\phi_{A_{i}}} & \cos{\theta_{A_{i}}} \\
    \sin{\theta_{A_{j}}}\cos{\phi_{A_{j}}} & \sin{\theta_{A_{j}}}\sin{\phi_{A_{j}}} & \cos{\theta_{A_{j}}} \\
    \alpha_{11} & \alpha_{12} & \alpha_{13} \\
  \end{array}\right|\frac{\partial N_{J}^{2}}{N_{J}^{2}\partial d_{A_{i}A_{j}}} , \nonumber \\
  & &
  \label{p1}
\end{eqnarray}
and
\begin{eqnarray}
& & d\gamma_{cB_{i}B_{j}} \nonumber \\
&=&\frac{dk}{2}
  \left|\begin{array}{ccc}
    \sin{\theta_{B_{i}}}\cos{\phi_{B_{i}}} & \sin{\theta_{B_{i}}}\sin{\phi_{B_{i}}} & \cos{\theta_{B_{i}}} \\
    \sin{\theta_{B_{j}}}\cos{\phi_{B_{j}}} & \sin{\theta_{B_{j}}}\sin{\phi_{B_{j}}} & \cos{\theta_{B_{j}}} \\
    \alpha_{21} & \alpha_{22} & \alpha_{23} \\
  \end{array}\right|\frac{\partial N_{J}^{2}}{N_{J}^{2}\partial d_{B_{i}B_{j}}} , \nonumber \\
& &  \label{p2}
\end{eqnarray}
where
\begin{eqnarray*}
  \alpha_{11} &=& (\cos{\theta_{A_{j}}}\cos{\phi_{A_{j}}}\theta_{A_{j}}^{'}-\sin{\theta_{A_{j}}}\sin{\phi_{A_{j}}}\phi_{A_{j}}^{'})-(j\rightarrow i), \\
  \alpha_{12} &=& (\cos{\theta_{A_{j}}}\sin{\phi_{A_{j}}}\theta_{A_{j}}^{'}+\sin{\theta_{A_{j}}}\cos{\phi_{A_{j}}}\phi_{A_{j}}^{'})-(j\rightarrow i), \\
  \alpha_{13} &=& -\sin\theta_{A_{j}}\theta_{A_{j}}^{'}-(j\rightarrow i),\\
  \alpha_{21} &=& (\cos{\theta_{B_{j}}}\cos{\phi_{B_{j}}}\theta_{B_{j}}^{'}-\sin{\theta_{B_{j}}}\sin{\phi_{B_{j}}}\phi_{B_{j}}^{'})-(j\rightarrow i), \\
  \alpha_{22} &=& (\cos{\theta_{B_{j}}}\sin{\phi_{B_{j}}}\theta_{B_{j}}^{'}+\sin{\theta_{B_{j}}}\cos{\phi_{B_{j}}}\phi_{B_{j}}^{'})-(j\rightarrow i), \\
  \alpha_{23} &=& -\sin\theta_{B_{j}}\theta_{B_{j}}^{'}-(j\rightarrow i) .
\end{eqnarray*}
Since $\theta_{A_{i}}=\pi-\theta_{B_{i}}$ and $\phi_{A_{i}}=\phi_{B_{i}}$, we have $d_{A_{i}A_{j}}=d_{B_{i}B_{j}}$, $\frac{\partial N_{J}^{2}}{N_{J}^{2}\partial d_{A_{i}A_{j}}}=\frac{\partial N_{J}^{2}}{N_{J}^{2}\partial d_{B_{i}B_{j}}}$, $\theta_{A_{i}}^{'}=-\theta_{B_{i}}^{'}$ and $\phi_{A_{i}}^{'}=\phi_{B_{i}}^{'}$, where $\theta^{'}=d \theta/dk$ and $\phi^{'}=d \phi/dk$. Substituting them into (\ref{p1}) and (\ref{p2}) we find that $d\gamma_{cA_{i}A_{j}}+d \gamma_{cB_{i}B_{j}}=0$. Similarly, we can derive $d\gamma_{cA_{i}B_{j}}+ d \gamma_{cB_{i}A_{j}}=0$, $d\gamma_{cA_{i}C_{l}}+d\gamma_{cB_{i}C_{l}}=0$.

For the correlation between $A_{i}$ and $B_{i}$, we know that $\theta_{A_{i}}=\pi-\theta_{B_{i}}=\theta$ and $\phi_{A_{i}}=\phi_{B_{i}}=\phi$, so
\begin{eqnarray}
& & d\gamma_{cA_{i}B_{i}} \nonumber \\
 &=& \frac{dk}{2}\left|
  \begin{array}{ccc}
    \sin\theta \cos\phi & \sin\theta \sin\phi & \cos\theta \\
    \sin\theta \cos\phi & \sin\theta \sin\phi & -\cos\theta \\
    0 & 0 & 2 \theta^{'} \sin\theta \\
  \end{array}
\right|\frac{\partial N_{J}^{2}}{N_{J}^{2}\partial d_{A_{i}B_{i}}} \nonumber \\
&=& 0 \nonumber.
\end{eqnarray}
At last, the correlation between $C_{l}$ and $C_{k}$ is always $0$, because they are on the same plane so that $\vec{u}_{l}\times\vec{u}_{k}\bullet d(\vec{u}_{l}-\vec{u}_{k})$ vanishes apparently. Hence, we have
\begin{eqnarray}
d\gamma_{c}&=&(d\gamma_{cA_{i}A_{j}}+d \gamma_{cB_{i}B_{j}}) + (d\gamma_{cA_{i}B_{j}}+d\gamma_{cB_{i}A_{j}}) + \nonumber \\
& & (d\gamma_{cA_{i}C_{l}}+d\gamma_{cB_{i}C_{l}}) + d\gamma_{cA_{i}B_{i}}+d\gamma_{cC_{l}C_{k}} \nonumber \\
&=& 0 \nonumber,
\end{eqnarray}
and the sum of correlation phases $\gamma_{c}$ is zero as the summation terms cancel out each other.



\begin{references}
\bibitem{Majorana1932} E. Majorana, Nuovo Cimento {\bf 9}, 43 (1932).

\bibitem{Schwinger} J. Schwinger, \emph{in Quantum theory of Angular Momentum}, edited by L. C. Biendenharn and H. Van Dam (Academic Press New York, 1965)

\bibitem{Hannay} J. H. Hannay, J. Phys. A {\bf 31}, L53 (1998).

\bibitem{Niu2012} Q. Niu, Physics {\bf 5}, 65 (2012).

\bibitem{Bruno2012} P. Bruno, Phys. Rev. Lett. \textbf{108}, 240402 (2012).

\bibitem{Fu2014} H. D. Liu and L. B. Fu, Phys. Rev. Lett. \textbf{113},  240403 (2014).

\bibitem{Bloch} F. Bloch and I. I. Rabi, Rev. Mod. Phys. {\bf 17}, 237 (1945).


\bibitem{N-qubit} H. M\"{a}kel\"{a} and A. Messina, Phys. Rev. A {\bf 81}, 012326 (2010); Phys. Scr. \textbf{T140}, 014054 (2010).

\bibitem{Ganczarek} W. Ganczarek, M. Ku\'{s}, and K. Zyczkowski, Phys. Rev. A
{\bf 85}, 032314 (2012).

\bibitem{Ribeiro} P. Ribeiro and R. Mosseri, Phys. Rev. Lett. {\bf 106}, 180502 (2011).

\bibitem{Markham}  D. J. H. Markham,  Phys. Rev. A {\bf 83}, 042332 (2011).

\bibitem{Mandilara} A. Mandilara, T. Coudreau, A. Keller, and P. Milman, Phys. Rev. A {\bf 90}, 050302(R) (2014).


\bibitem{Vidal} P. Ribeiro, J. Vidal, and R. Mosseri, Phys. Rev. Lett. {\bf 99}, 050402 (2007); Phys. Rev. E 78, 021106 (2008).

\bibitem{Demler} R. Barnett, A. Turner, and E. Demler, Phys. Rev. Lett. {\bf 97}, 180412
(2006); Phys. Rev. A {\bf 76}, 013605 (2007).

\bibitem{Lamacraft} A. Lamacraft, Phys. Rev. B {\bf 81}, 184526 (2010).

\bibitem{Ueda}D. Stamper-Kurn and M. Ueda, Rev. Mod. Phys. {\bf 85}, 1191 (2013); Y. Kawaguchi and M. Ueda, Phys. Rep. {\bf 520}, 253 (2012).

\bibitem{Zhaihui} B. Lian, T. L. Ho, and H. Zhai, Phys. Rev. A {\bf 85},
051606(R) (2012); X. L. Cui, B. Lian, T. L. Ho, B. L. Lev, and H. Zhai,
Phys. Rev. A, \textbf{88}, 011601(R) (2013).

\bibitem{TI-RMP} M. Z. Hasan and C. L. Kane, Rev. Mod. Phys. \textbf{82}, 3045
(2010); X.-L. Qi and S.-C. Zhang,  Rev. Mod. Phys. \textbf{83},
1057 (2011).

\bibitem{SSH}
W. P. Su, J. R. Schrieffer, and A. J. Heeger, , Phys. Rev. Lett. {\bf 42},
1698 (1979); Phys. Rev. B {\bf 22}, 2099 (1980).

\bibitem{Ryu} S. Ryu and Y. Hatsugai, Phys. Rev. Lett. {\bf 89}, 077002
(2002).

\bibitem{Montambaux} P. Delplace, D. Ullmo, and G. Montambaux, Phys. Rev. B {\bf 84},
195452 (2011).



\bibitem{Shen-book} S.-Q. Shen, \emph{Topological Insulators} (Springer-Verlag, Hei-
delberg, 2013).


\bibitem{Berry} M. V. Berry, Roc. R. Soc. A {\bf 392} 45 (1984).

\bibitem{Zak} J. Zak, Phys. Rev. Lett. {\bf 62} 2747 (1989).

\bibitem{Niu-RMP} D. Xiao, M.C. Chang and Q. Niu, Rev. Mod. Phys. \textbf{82},
1959 (2010).

\bibitem{cold1} M. Atala, M. Aidelsburger, J. T. Barreiro, D. Abanin, T. Kitagawa, E. Demler and I. Bloch, Nature Phys. {\bf 9}, 795 (2013).


\bibitem{SS} W. P. Su and J. R. Schrieffer, Phys. Rev. Lett. {\bf 46}, 738 (1981).


\bibitem{Ganeshan} S. Ganeshan, K. Sun, S.D. Sarma, Phys. Rev. Lett. {\bf 110} 180403 (2013).

\bibitem{Guo} H. M. Guo and S. Chen, Phys. Rev. B {\bf 91}, 041402(R) (2015).

\bibitem{inversion} T. L. Hughes, E. Prodan
and B. A. Bernevig, Phys. Rev. B {\bf 83}, 245132 (2011); C.-K Chiu, H. Yao, and S. Ryu, Phys. Rev. B {\bf 88}, 075142
(2013); Y.-M. Lu and D.-H. Lee, arXiv: 1403.5558.


\end{references}
\end{document}